\documentclass[prd, twocolumn, nofootinbib, superscriptaddress]{revtex4-2}

\usepackage{aas_macros}
\usepackage{graphicx}
\usepackage[caption=false]{subfig}
\usepackage{siunitx}
\usepackage{mathrsfs}
\usepackage{amsmath,amssymb}
\usepackage{bm}
\usepackage{braket}
\usepackage{listings}
\usepackage{cases}
\usepackage{here}
\usepackage{comment}
\usepackage{soul}
\usepackage{cancel}
\usepackage{cases}
\usepackage[utf8]{inputenc}
\usepackage{url}
\usepackage{longtable}
\usepackage[normalem]{ulem}
\usepackage{xspace}
\usepackage[colorlinks=true
,urlcolor=blue
,anchorcolor=blue
,citecolor=blue
,filecolor=blue
,linkcolor=blue
,menucolor=blue
,linktocpage=true
,pdfproducer=medialab
,pdfa=true
]{hyperref}
\usepackage{animate}

\begin{document}

% \preprint{APS/123-QED}

\title{Galaxy-galaxy strong lensing cross-section with fuzzy dark matter model}

\author{Hiroki Kawai}
\email{hiroki.kawai@phys.s.u-tokyo.ac.jp}
\affiliation{Department of Physics, The University of Tokyo, Bunkyo, Tokyo 113-0033, Japan}
\affiliation{Center for Frontier Science, Chiba University, 1-33 Yayoicho, Inage, Chiba 263-8522, Japan}
\affiliation{INAF-Osservatorio di Astrofisica e Scienza dello Spazio di Bologna, Via Piero Gobetti 93/3, 40129 Bologna, Italy}

\author{Massimo Meneghetti}
\affiliation{INAF-Osservatorio di Astrofisica e Scienza dello Spazio di Bologna, Via Piero Gobetti 93/3, 40129 Bologna, Italy}
\affiliation{Istituto Nazionale di Fisica Nucleare, viale Berti Pichat 6/2, 40127 Bologna, Italy}

\date{\today}

\begin{abstract}
The galaxy-galaxy strong lensing (GGSL) cross-section in observed galaxy clusters has been reported to be more than an order of magnitude higher than the theoretical prediction by the standard cold dark matter (CDM) model.
In this study, we focus on the fuzzy dark matter (FDM) model and study the GGSL cross-section numerically and analytically.
We find that FDM subhalos can produce larger cross-sections than the CDM subhalos due to the presence of the soliton core.
The maximum cross-section is obtained when the core radius is about the same as the size of the critical curve.
The peak ratio of the cross-sections between the FDM subhalos and the CDM subhalos is about two when including the baryon distribution, indicating that the FDM with any masses might not produce the expected observed cross-section.
\end{abstract}


\maketitle

\section{Introduction} \label{sec:intro}
% Small-scale problems → GGSL probability discrepancy
While the standard model of the universe, the so-called $\Lambda$ Cold Dark Matter ($\Lambda$CDM) model, which assumes that the dark matter is non-relativistic and interacts only through gravity, can explain a wide range of observations at large scales ({\it{e.g.,}} \citep{2004ApJ...606..702T, 2019MNRAS.489.2247C}), there are discrepancies between the $\Lambda$CDM predictions and the observations at small scales, known as small-scale problems (see \citep{2017ARA&A..55..343B} for a recent review).
One of these problems identified recently is the discrepancy of the galaxy-galaxy strong lensing (GGSL) cross-section in galaxy clusters \citep{2020Sci...369.1347M, 2022A&A...668A.188M, 2023A&A...678L...2M}, which we focus on in this paper. 
The GGSL cross-section is defined by the total area enclosed by the secondary caustics on the source plane, produced by substructures such as subhalos and galaxies.
% , divided by the full field of view of the lens plane projected onto the source plane.
The observed cross-section is an order of magnitude higher than the prediction by hydrodynamical $\Lambda$CDM simulations, which might indicate that the observed mass distribution is more concentrated than the CDM subhalos.
Note that this trend is the opposite of other small-scale problems such as the core-cusp problem \citep{1994Natur.370..629M} and the missing satellite problem \citep{1999ApJ...524L..19M} since they suffer from the overdense mass distribution of the CDM.

% Possible solutions
The general idea to alleviate these small-scale problems is to consider the baryon physics and/or alternative dark matter models, which would change the mass distribution inside halos and subhalos.
Several studies have already been conducted on whether they can resolve the discrepancy in the GGSL cross-section.
The effect of the baryon physics on the cross-section is studied by hydrodynamical simulations conducted by \citet{2021MNRAS.505.1458B} and \citet{2021MNRAS.504L...7R}.
They show that subhalos with mass $M_{\rm sh} \gtrsim 10^{11}\ M_{\odot}$ can be more compact to produce the higher cross-section with low active galactic nuclei (AGN) efficiencies.
However, the number of stellar components in these subhalos exceeds one found in observations with such a small AGN efficiency, indicating that the discrepancy cannot be solved solely by AGN physics \citep{2022A&A...665A..16R}.
\citet{2024arXiv240416951T} consider the mass redistribution by changing the concentration and the tidal radius of the CDM subhalos since the observation only constrains the internal total mass, showing that there still exists the tension within the CDM paradigm.
\citet{2024arXiv240617024D} study the GGSL cross-section with the self-interacting dark matter (SIDM) model \citep{2000PhRvL..84.3760S}.
They show that the inner density profile of $\rho \propto r^{-\gamma}$ with the power-law index $\gamma > 2.5$ can alleviate the discrepancy, which the core-collapsed SIDM subhalos can achieve.
However, we still need to explore alternative solutions, since they do not consider the details such as a fraction of the core-collapsed objects and the SIDM subhalo mass function.
Further detailed studies on the GGSL cross-section with the SIDM model might be possible with a recent simulation \citep{2024A&A...687A.270R}.

% FDM model + constraint
As another alternative to CDM, we focus on the fuzzy dark matter (FDM) model \citep{2000PhRvL..85.1158H} and study the GGSL cross-section.
The FDM is a scalar particle whose mass is around $m \simeq 10^{-24} - 10^{-20}\ {\rm eV}/c^{2}$ and a specific candidate for the FDM in particle physics is the axion-like particle appearing in the string theory \citep{2017PhRvD..95d3541H}.
The de Broglie wavelength on the scale of the universe has the potential to solve the small-scale problems because of its wave nature.
One of the distinguishable features of FDM halos is the soliton core at the center surrounded by the Navarro-Freck-White (NFW) profile, as shown by the cosmological FDM simulations which solve the Schr{\"o}dinger-Poisson (SP) equation \citep{2014NatPh..10..496S, 2018PhRvD..97h3519M, 2021MNRAS.506.2603M}.
The soliton core corresponds to the lowest energy state and its density profile can be expressed by the ground state solution of the SP equation.
The soliton core has two parameters, namely the soliton core mass (or core radius) and the FDM mass. 
The core mass is related to the halo mass via the core-halo mass relation \citep{2014PhRvL.113z1302S, 2022MNRAS.511..943C, 2022PhRvD.106j3532T, 2024PhRvD.110b3519K}.
For a fixed core mass, the central core density increases and the core radius decreases as the FDM mass increases.
Therefore, it is expected that the GGSL cross-section would depend on the FDM mass, which is the motivation for this study.
By investigating this relationship, we aim to determine whether the FDM model can produce a sufficiently large GGSL cross-section and to identify the preferred mass for FDM particles.

% This work
In the present paper, we first construct an analytic model of the cross-section of a single FDM subhalo in the host halo.
We study the dependence of the cross-section on the FDM mass, the subhalo mass, and the distance from the center of the host halo.
We then calculate the total cross-section by integrating the product of the subhalo mass function and the cross-section by a single subhalo.
We also consider the baryon distribution assuming the stellar-to-halo mass relation and study how the baryon distribution affects the cross-section. 

% Section
In this paper, we first review the FDM model in Sec.~\ref{sec:fdm_basic}.
We then show our analytic model of the GGSL cross-sections of FDM subhalos in Sec.~\ref{sec:ggsl_fdm}.
We then study the effect of the baryon distribution and compare our analytic estimations with the observational results in Sec.~\ref{sec:compare_obs}.
We finally show the summary and discussions in Sec.~\ref{sec:summary_discussion}.
Throughout the present paper, the halo size is determined by the virial radius, and the concentration-halo mass relation for the CDM halos is determined according to the simulation results \citep{2021MNRAS.506.4210I}.
We use cosmological parameters of $\Omega_{\rm m0}=0.3111$, $\Omega_{\rm b0} = 0.0490$, and $H_{0} = 67.66\ {\rm km/s/Mpc}$, which are obtained from the Planck satellite \citep{2020A&A...641A...6P}.

\section{Fuzzy dark matter} \label{sec:fdm_basic}
% de Broglie wavelength
The FDM is a scalar particle minimally coupled to gravity without self-interaction whose mass is around $m \simeq 10^{-24}-10^{-20}\ {\rm eV}/c^{2}$.
Due to the small mass of FDM, the corresponding de Broglie wavelength becomes a cosmological scale,
\begin{equation}
    \frac{\lambda_{\rm dB}}{2\pi} = \frac{\hbar}{mv} = 1.92\ {\rm kpc} \left(\frac{mc^{2}}{10^{-22}\ {\rm eV}}\right)^{-1} \left(\frac{v}{10\ {\rm km/s}}\right)^{-1}, \label{fdm_dB_wavelength}
\end{equation}
with $\hbar$ and $v$ being the Dirac constant and the velocity, respectively.
Since the wave nature is dominant below the scale of the de Broglie wavelength, the mass distribution differs from that of the standard CDM universe.
This is particularly evident in the density profiles of (sub)halos and their overall distribution, namely, the (sub)halo mass function.

\subsection{Density profile} \label{subsec:fdm_density_profile}
% FDM density profile (soliton core)
The FDM halo has the soliton core at the center surrounded by the NFW density profile.
The density profile of the soliton core can be expressed by \cite{2014NatPh..10..496S}
\begin{equation}
    \rho_{\rm sol}(r) = \frac{\rho_{\rm c}}{\{1+0.091(r/r_{\rm c})^{2}\}^{8}}, \label{soliton_emp}
\end{equation}
where the central density $\rho_{\rm c}$ is
\begin{equation}
    \rho_{\rm c} = 0.019 \left(\frac{mc^{2}}{10^{-22}\ {\rm eV}}\right)^{-2} \left(\frac{r_{\rm c}}{\rm kpc}\right)^{-4}\ M_{\odot}\ \rm{pc}^{-3}. \label{rho_c_dep}
\end{equation}
Here $r_{\rm c}$ represents the core radius where the density drops to half of the central density.
The core mass is defined as the enclosed mass within the core radius,
\begin{eqnarray}
    M_{\rm c} &=& M_{\rm sol} (< r_{\rm c}) \nonumber \\
    &=& 5.3 \times 10^{7} \left(\frac{mc^{2}}{10^{-22}\ {\rm eV}}\right)^{-2} \left(\frac{r_{\rm c}}{{\rm kpc}}\right)^{-1}\ M_{\odot} \label{rctrue_Mc}.
\end{eqnarray}
As can be seen from the relations above, the soliton core profile has two parameters, namely, the core mass (or core radius) and the FDM mass.

% FDM density profile (NFW profile, c-M relation)
The outer NFW profile can be written in the form of \cite{1997ApJ...490..493N}
\begin{equation}
    \rho_{\rm NFW}(r) = \frac{\rho_{\rm s}}{(r/r_{\rm s})(1+r/r_{\rm s})^{2}}, \label{nfw}
\end{equation}
where $r_{\rm s}$ is the scale radius and $\rho_{\rm s}$ is the scale density.
Instead of using the scale radius and density, the halo mass $M_{\rm h}$ and the concentration parameter $c_{\rm vir}$ are often used as the two parameters to determine the NFW profile.
The concentration parameter is related to the halo mass through the concentration-halo mass relation, $c_{\rm vir}$-$M_{\rm h}$ relation ({\it{e.g.}}, \citep{2014ApJ...797...34M, 2021MNRAS.506.4210I} for CDM halos). 
The concentration of FDM halos shows the suppression to that of CDM halos below the half-mode mass as shown by \citep{2022MNRAS.515.5646D, 2022MNRAS.517.1867L},
\begin{equation}
    c_{\rm vir}(M_{\rm h}, z; {\rm FDM}) = c_{\rm vir}(M_{\rm h}, z; {\rm CDM})\ F\left(\frac{M_{\rm h}}{M_{\rm h}^{\rm hm}}\right), \label{cfdm_def}
\end{equation}
where the function $F$ is $F(x) = (1+ax^{b})^{c}$ and the half-mode mass $M_{\rm h}^{\rm hm}$ is 
\begin{equation}
    M_{\rm h}^{\rm hm} = 3.8\times 10^{10}\ M_{\odot} \left(\frac{mc^{2}}{10^{-22}\ \rm{eV}}\right)^{-\frac{4}{3}} \label{half_mode_mass}.
\end{equation}
In this paper, we set the parameters $(a,b,c) = (9.431, -1.175, -0.232)$, following the result by \citep{2022MNRAS.515.5646D, 2024PhRvD.110b3519K}.
Note that the $c_{\rm vir}$-$M_{\rm h}$ relation of FDM halos shows the turnover around $M_{\rm h} \simeq 4 M_{\rm h}^{\rm hm} (\equiv M_{\rm h}^{\rm 4hm})$, four times larger than the half-mode mass.

% FDM density profile (core-halo mass relation, density profile)
The relation between the core mass and halo mass, the core-halo mass relation (CHMR), determines the total density profile of FDM halos for a given FDM mass.
The CHMR is studied numerically and analytically \citep{2014PhRvL.113z1302S, 2022MNRAS.511..943C, 2022PhRvD.106j3532T, 2024PhRvD.110b3519K}.
In this study, we follow the recent study conducted by \citet{2024PhRvD.110b3519K}, where the core profile is assumed to be generated by the mass redistribution of the NFW profile within the characteristic radius calculated from the relaxation time condition, and the CHMR at redshift $z < 1$ is expressed in the form of
\begin{equation}
    M_{\rm c} = 
    \begin{cases}
        M_{\rm c}^{\rm min} \left(\frac{M_{\rm h}}{M_{\rm h}^{\rm min}}\right)^{0.32}\ (M_{\rm h}^{\rm min} < M_{\rm h} < M_{\rm h}^{\rm 4hm}) \\
        M_{\rm c}^{\rm 4hm} \left(\frac{M_{\rm h}}{M_{\rm h}^{\rm 4hm}}\right)^{0.07} (M_{\rm h} > M_\mathrm{h}^{\rm 4hm})
    \end{cases}, \label{analytic_expression}
\end{equation}
where $M_{\rm h}^{\rm min}$ and $M_{\rm c}^{\rm min}$ denote the minimum halo and core masses, respectively.
The minimum halo mass is 
\begin{align}
    M_{\rm h}^{\rm min} & \simeq 4 \times 10^{7}\ M_{\odot} \left(\frac{mc^{2}}{10^{-22}\ {\rm eV}}\right)^{-\frac{3}{2}} \nonumber \\
    & \quad \times \left(\frac{\rho_{\rm m0}}{40.8\ M_{\odot}\ {\rm kpc^{-3}}}\right)^{\frac{1}{2}} \left(\frac{t_{0}}{13.8\ {\rm Gyr}}\right)^{\frac{1}{2}}, \label{min_mass}
\end{align}
and $M_{\rm c}^{\rm min} \simeq 0.34 M_{\rm h}^{\rm min}$.
Here $\rho_{\rm m,0}$ and $t_{0}$ are the matter density and the age of the Universe at $z=0$, respectively. 
The two formulas in Eq.~\eqref{analytic_expression} are continuous, that is, $M_{\rm c}^{\rm 4hm}$ is defined as the core mass when the halo mass equals to $M_{\rm h}^{\rm 4hm}$.

\begin{figure}
    \includegraphics[width=\columnwidth]{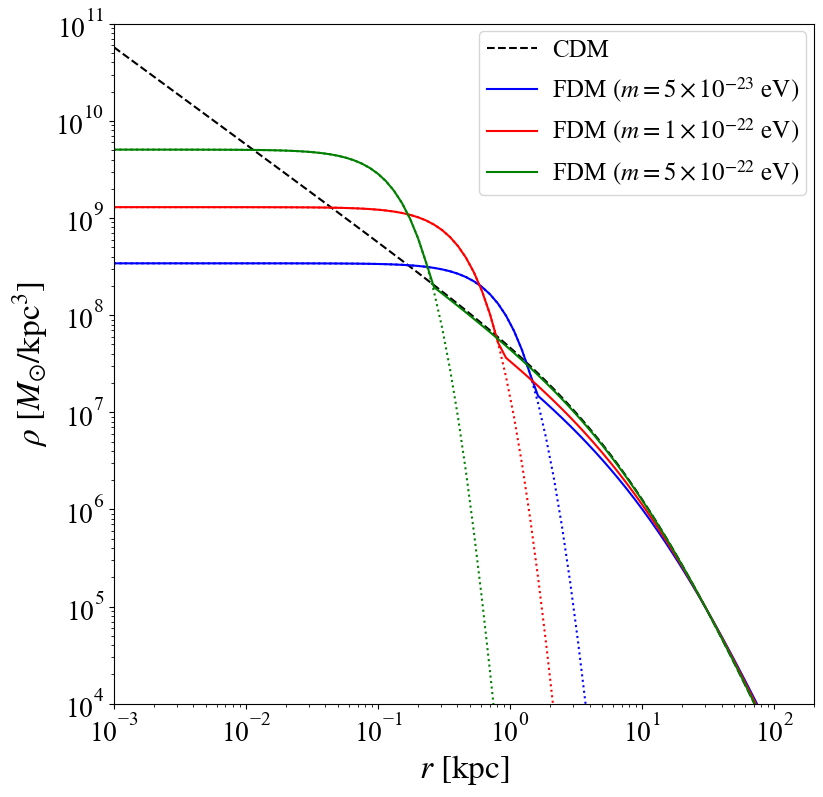}
    \caption{The comparison between the CDM and the FDM density profiles with different FDM masses.
    The black dashed line is the NFW profile in the CDM halo.
    The blue, red, and green solid and dotted lines show the total density profiles of the FDM halo and their soliton core profiles with $m = 5 \times 10^{-23} {\rm eV}/c^{2}$, $1 \times 10^{-22} {\rm eV}/c^{2}$, and $5 \times 10^{-22} {\rm eV}/c^{2}$ respectively.
    Here, we set $M_{\rm h} = 10^{11}\ M_{\odot}$ and $z = 0$.
    We use the $c_{\rm vir}$-$M_{\rm h}$ relation to obtain the concentration of the CDM halo and multiply the suppression function $F$ to obtain those of the FDM halos.
    }
    \label{fig:fdm_density_profile}
\end{figure}

Using the CHMR above, the total density profile can be expressed in the form of
\begin{equation}
    \rho(r) = 
    \begin{cases}
        \rho_{\rm sol}(r)\ \ \ \ \ \ \ \ \ \ \ \ \ \ \ \ \ \ \ \ \ \ \ \ (r < r_{\rm c}/2) \\
        {\rm max}[\rho_{\rm sol}(r), \rho_{\rm NFW}(r)]\ \ \ \ (r > r_{\rm c}/2)
    \end{cases}. \label{total_rho}    
\end{equation}
In Fig.~\ref{fig:fdm_density_profile}, we show the density profile of the FDM halo with different FDM masses.
The FDM density profile exceeds the standard CDM profile near the core radius due to the presence of the soliton core. 
As the FDM particle mass increases, the core radius shrinks, and the central density becomes larger. 
These changes affect the size of the critical curve, and consequently, the cross-section.

\subsection{Distribution of subhalos} \label{subsec:subhalo_distribution}
% CDM subhalo mass function
The dark matter nature changes the mass distribution of the compact objects. 
Here we review the distribution of the subhalos inside halos, known as the subhalo mass function. 
The CDM subhalo mass function has been extensively studied numerically and analytically ({\it{e.g.}}, \citep{2004ApJ...604L..73L, 2004MNRAS.355..120O, 2005MNRAS.359.1029V, 2008MNRAS.386.2135G, 2008MNRAS.387..689G, 2016MNRAS.457.1208H, 2016MNRAS.458.2848J, 2018PhRvD..97l3002H, 2019Galax...7...68A, 2020ApJ...901...58O, 2022MNRAS.517.2728H}).
This study uses the semianalytic model of the CDM subhalo mass function obtained by \citet{2022MNRAS.517.2728H}.
In this model, the evolved mass function, which is related to the observables in the current Universe, is calculated by the combination of the unevolved mass function and the tidal evolution model by \citet{2018PhRvD..97l3002H}.
Here, they assume that the increase of the host halo mass is solely determined by accretions of smaller halos, and they calculate the unevolved mass function by integrating all the contributions of accreting halos.
We use the open source code \footnote{\url{https://github.com/shinichiroando/sashimi-c}} provided by \citet{2022MNRAS.517.2728H} to calculate the subhalo mass function.

% projected CDM subhalo mass function
Assuming that the spatial distribution of subhalos follows the density distribution of the host halo, which is supported by the numerical simulation \citep{2017MNRAS.472..657J}, the surface number density of subhalos becomes proportional to the surface density profile of the host halo,
\begin{equation}
    \frac{dN_{\rm sh}}{dS} = \frac{N_{\rm sh} \Sigma(d_{\rm sh};M_{\rm hh})}{M_{\rm hh}},
\end{equation}
where $d_{\rm sh}$ is the distance to the host halo center and $dS = 2\pi d_{\rm sh} dd_{\rm sh}$ assuming the spherical symmetry.
The projected subhalo mass function can be expressed as
\begin{equation}
    \frac{d^{2}N_{\rm sh}}{dM_{\rm sh} dS} = \frac{dN_{\rm sh}}{dM_{\rm sh}} \frac{\Sigma(d_{\rm sh};M_{\rm hh})}{M_{\rm hh}},
\end{equation}
where $M_{\rm sh}$ denotes the subhalo mass.
In this study, the density profile for the host halo is set to the NFW profile.
The subhalo mass function $dN_{\rm sh}/dM_{\rm sh}$ is calculated by using the semianalytical model by \citet{2022MNRAS.517.2728H} as stated above.

\begin{figure}
    \includegraphics[width=\columnwidth]{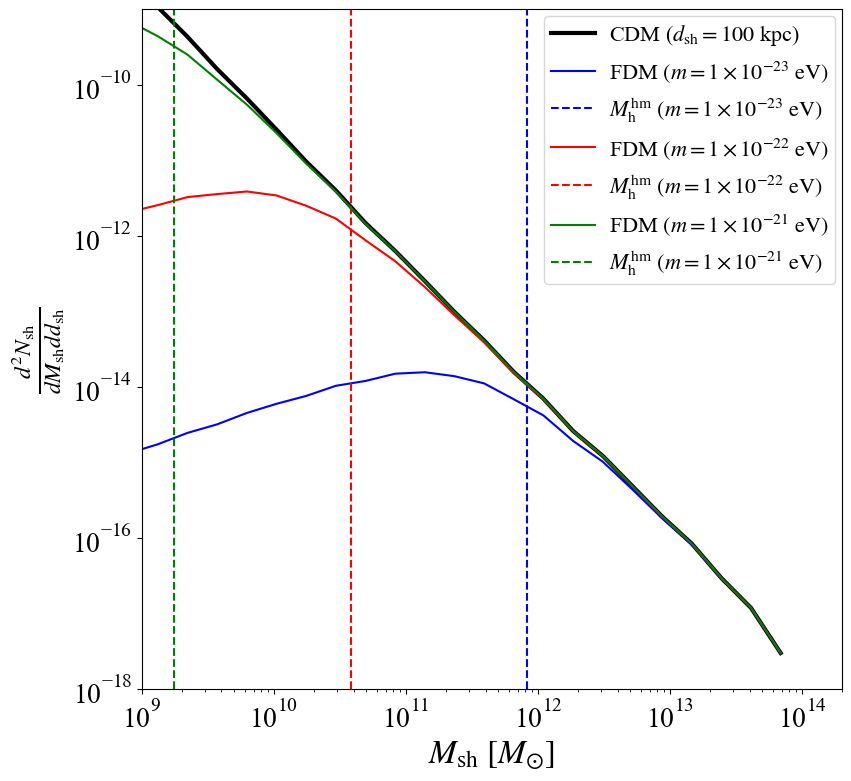}
    \caption{The evolved projected subhalo mass function in the case of CDM and FDM with different FDM masses.
    Here we assume that the host halo mass is $M_{\rm hh} = 10^{15}\ M_{\odot}$ and the density profile is set to the NFW profile.
    We use the concentration-halo mass relation obtained by \citet{2021MNRAS.506.4210I}.
    The position of the subhalos is set to $d_{\rm sh} = 100\ {\rm kpc}$ and we use the relation $dS = 2\pi d_{\rm sh} dd_{\rm sh}$.
    The black solid line shows the CDM case calculated by \citep{2022MNRAS.517.2728H} with their provided code.
    The blue, red, and green solid lines are the FDM cases with FDM masses $m = 1\times 10^{-23}\ {\rm eV}/c^{2}$, $1\times 10^{-22}\ {\rm eV}/c^{2}$, and $1\times 10^{-21}\ {\rm eV}/c^{2}$, respectively.
    The suppression of the subhalo mass function below the half-mode masses, shown in the vertical dashed lines, is calculated by the function $F$ with $(a,b,c)=(0.36, -1.1, -2.2)$.
    }
    \label{fig:project_sbmf_fdm}
\end{figure}

Since the linear power spectrum is suppressed due to the wave nature of the FDM, the abundance of the subhalos is reduced compared to the CDM case.
The suppression occurs below the half-mode mass as shown by \citep{2016ApJ...818...89S, 2022MNRAS.517.1867L},
\begin{equation}
    \left.\frac{d^{2}N}{dM_{\rm sh}dS}\right|_{\rm FDM} = \left.\frac{d^{2}N}{dM_{\rm sh}dS}\right|_{\rm CDM}\ F\left(\frac{M_{\rm sh}}{M_{\rm h}^{\rm hm}}\right), \label{halo_massfunc_fdm}
\end{equation}
where the suppression function $F$ is the same as Eq.~\eqref{cfdm_def}, but now $(a,b,c)=(0.36, -1.1, -2.2)$ \citep{2022MNRAS.517.1867L}.
In Fig.~\ref{fig:project_sbmf_fdm}, we show the evolved projected subhalo mass function in the case of CDM and FDM.
As the FDM mass decreases, the turnover occurs at a higher subhalo mass due to the larger de Broglie wavelength.

\section{GGSL cross-section with FDM subhalos} \label{sec:ggsl_fdm}
In this section, we show our modeling of the GGSL cross-section with FDM subhalos.
Here, we set the total host halo mass as $M_{\rm hh} = 10^{15}\ M_{\odot}$, and the density profile of the host halo is set to the NFW density profile.
Note that since the core radius of the host halo is small enough to be considered, we ignore the impact of the soliton core of the host halo.
We focus on the spherical host and subhalos for simplicity.
The redshifts of the source and lens planes are set to $z_{\rm s}=2$ and $z_{\rm l}=0.5$, respectively.
In these setups, the size of the critical curve of the host halo is about 14 arcsec.
First, we show the modeling of the GGSL cross-section of a single FDM subhalo in Sec.~\ref{subsec:ggsl_fdm_single}.
Considering the subhalo mass function, we calculate the total cross-section from all FDM subhalos, shown in Sec.~\ref{subsec:ggsl_fdm_total}.

\subsection{Single FDM subhalo} \label{subsec:ggsl_fdm_single}
In this subsection, we consider the cross-section of a single FDM subhalo within the host halo.
To obtain the cross-section, we conduct a numerical calculation by using the same code as used in \citet{2020Sci...369.1347M}.
Here, we first generate convergence maps (projected density fields) and calculate the deflection angle using ray-tracing simulations. 
By adding the deflection angle contributed by the host halo, we calculate the critical curve around the subhalos.

\begin{figure*}
    \includegraphics[width=\linewidth]{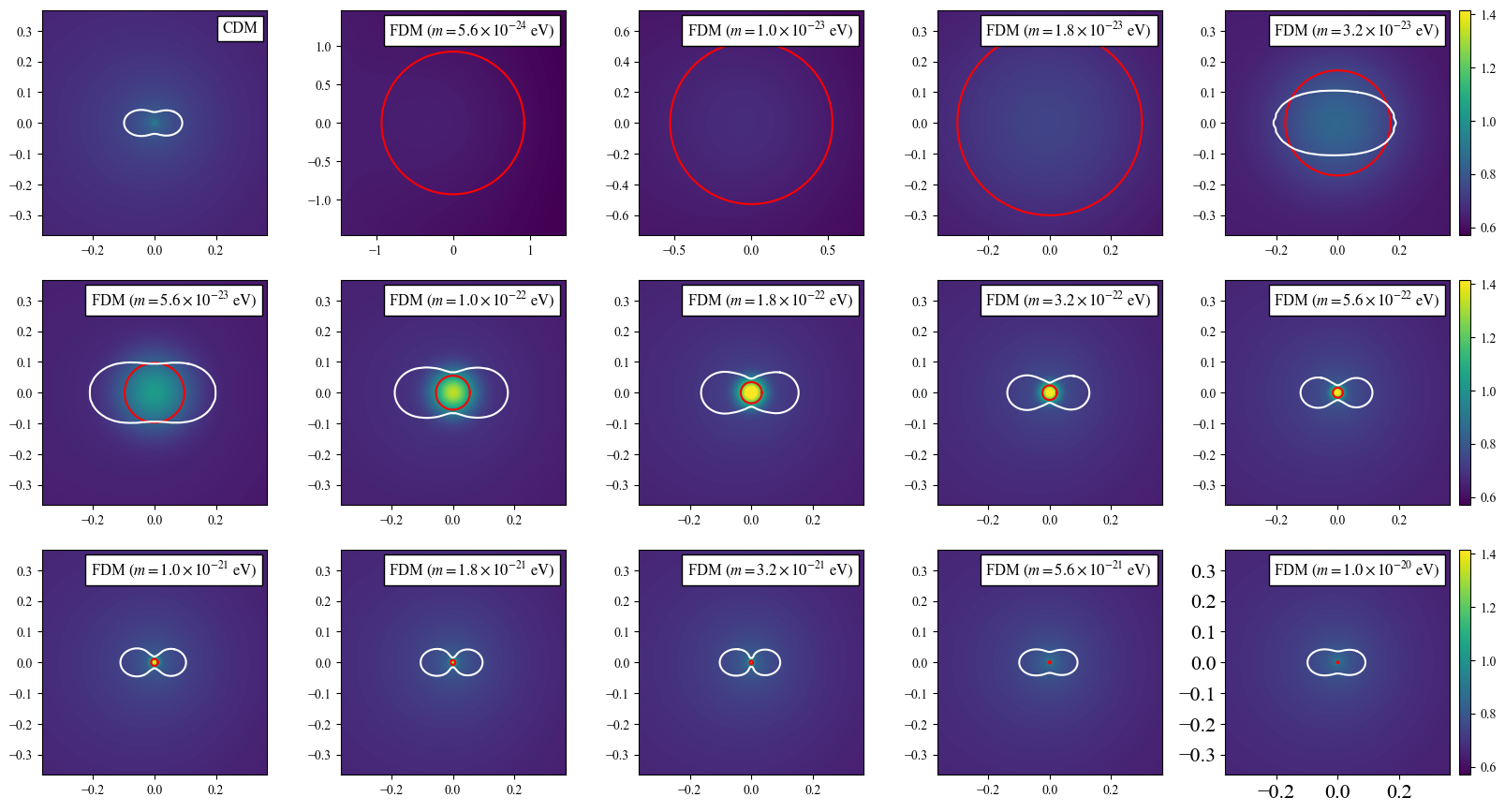}
    \caption{The convergence maps around the position of the subhalos with the CDM model and the FDM model with different FDM masses.
    The subhalos are located at 20 arcsec from the center of the host halo, $(x,y) = (-20,0)$, where the unit is arcsec.
    The masses of host halo and subhalos are set to $10^{15}\ M_{\odot}$ and $10^{11}\ M_{\odot}$, respectively.
    The redshifts of the source plane and lens plane are set to $z_{\rm s} = 2.0$ and $z_{\rm l} = 0.5$, respectively.
    The white lines show the secondary critical curves and the red lines show the core radius in the case of the FDM model.
    Note that the box sizes are different in the case with the FDM mass $m = 5.6\times 10^{-24}\ {\rm eV}/c^{2}$ and $1.0\times 10^{-23}\ {\rm eV}/c^{2}$.
    }
    \label{fig:massMap}
\end{figure*}

\begin{figure}
    \includegraphics[width=\columnwidth]{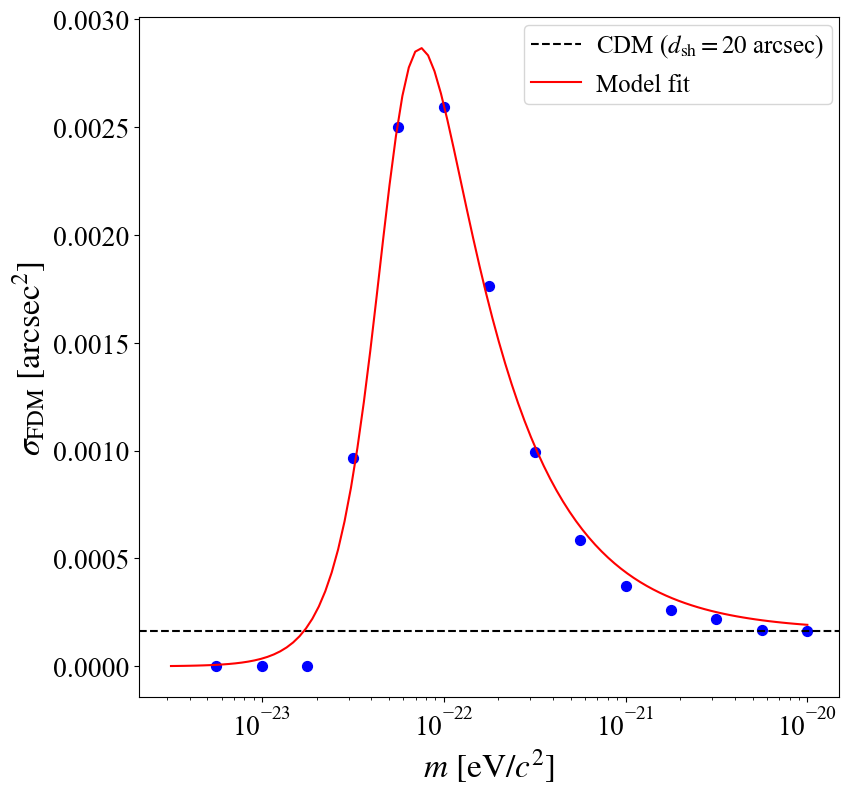}
    \caption{The GGSL cross-section of a single FDM subhalo in a cluster host as a function of the FDM mass.
    The parameters are the same as Fig.~\ref{fig:massMap}.
    The horizontal dashed line indicates the cross-section of the CDM subhalo.
    The red line shows the fitting result of our analytic model with the fitting parameters being $m_{\rm peak} = 10^{-22.3}\ {\rm eV}/c^{2}$, $\sigma_{\rm FDM}^{\rm peak} = 0.005$, and $\Delta_{\log_{10} m} = 0.23$.
    }
    \label{fig:ggsl_cs}
\end{figure}

% Mass map
In Fig.~\ref{fig:massMap}, we show the convergence map of the CDM subhalo and FDM subhalos with different FDM masses located at 20 arcsec from the center of the host halo.
We can see that the size of the core radii (red lines) becomes smaller as the FDM mass becomes larger as shown in Sec.~\ref{sec:fdm_basic}.
The secondary critical curves are shown in white lines.
As expected, the size of the critical curve depends on the FDM mass.
In Fig.~\ref{fig:ggsl_cs}, we plot the corresponding GGSL cross-sections.
We find that the cross-section is zero with sufficient small FDM masses because the central density of the soliton core is not dense enough to create the critical curve, {\it{i.e.}}, the convergence is less than one at the center.
In the limit of the large FDM masses, the cross-section is the same as that with the CDM subhalo since the effect of the soliton cores is negligible.
Notably, when the FDM masses are in the middle of the above regions and the core radii are around the same as the size of the critical curves, the cross-section becomes larger than the CDM case.
Since the peak cross-section is an order of magnitude larger than that of the CDM subhalo, it might indicate that the GGSL discrepancy is alleviated by considering the FDM model.

% Analytic model
We find that the following analytic formula well describes the GGSL cross-section of the single FDM subhalo,
\begin{eqnarray}
    && \sigma_{\rm FDM}(m; M_{\rm sh}, d_{\rm sh}) \nonumber \\
    && \hspace{3mm} = \frac{1}{2} \left\{1+\tanh\left(\frac{\log_{10}m - \log_{10}m_{\rm peak}}{\Delta_{\log_{10}m}}\right) \right\} \nonumber \\
    && \hspace{6mm} \times \left\{\sigma_{\rm FDM}^{\rm peak} \left(\frac{m}{m_{\rm peak}}\right)^{-1} + \sigma_{\rm CDM}(M_{\rm sh},d_{\rm sh})\right\}. \label{ggsl_cs_fdm_model}
\end{eqnarray}
Here we denote the peak cross-section by $\sigma_{\rm FDM}^{\rm peak}$, which can be seen in Fig.~\ref{fig:ggsl_cs}, and the corresponding FDM mass is denoted by $m_{\rm peak}$.
The first term indicates that the cross-section becomes zero below the peak FDM mass due to a shallow core density profile. 
The second term indicates that the cross-section falls in proportion to the inverse of FDM mass above the peak FDM mass, and approaches to the cross-section of the CDM subhalo.
When the FDM mass is above the peak mass, $m \gtrsim m_{\rm peak}$, the core radius is smaller than the Einstein radius, $x_{\rm c} \lesssim x_{\rm Ein}$, where $x$ denotes the two-dimensional coordinates on the projected field.
Considering the relation $\sigma_{\rm FDM} \propto x_{\rm Ein}^{2} \propto M(<x_{\rm Ein}) \simeq M_{\rm c} + M_{\rm NFW}(<x_{\rm Ein})$, where $M(<x)$ denotes the cylinder mass and $x_{\rm Ein}$ denotes the Einstein radius, and the core mass dependence on the FDM mass $M_{\rm c} \propto m^{-1}$, we can understand the dependence of the cross-section on the FDM mass in Eq.~\eqref{ggsl_cs_fdm_model}.
In Fig.~\ref{fig:ggsl_cs}, we show that our analytic formula, shown in a red solid line, well describes the behavior of the numerical results.

% Dependence on two parameters
The peak cross-section and the peak FDM mass depend on two parameters, $M_{\rm sh}$ and $d_{\rm sh}$.
As indicated from Fig.~\ref{fig:massMap}, the peak cross-section can be obtained where the core radius is around the same as the size of the critical curve (or approximately the Einstein radius), $x_{\rm c} \simeq x_{\rm Ein}$.
In this case, the cylinder mass within the critical curve can be approximated by the core mass, $x_{\rm Ein} \propto M_{\rm c}^{1/2}$, and by using the relation $x_{\rm c} = r_{\rm c} \propto M_{\rm sh}^{-1/3} m^{-1}$ and $M_{\rm c} \propto M_{\rm sh}^{1/3} m^{-1}$, we can obtain the peak FDM mass dependence on the parameters,
\begin{equation}
    m_{\rm peak}(M_{\rm sh},d_{\rm sh}) \propto M_{\rm sh}^{-1}.
\end{equation}
Note that the size of the critical curve is influenced by the contribution of the host halo and depends on the position of the subhalo.
While this could affect the peak FDM mass, numerical results show negligible dependence.
The dependence of the peak cross-section on the subhalo mass is now estimated as $\sigma_{\rm FDM}^{\rm peak} \propto x_{\rm c, peak}^{2} \propto M_{\rm sh}^{-2/3} m_{\rm peak}^{-2} \propto M_{\rm sh}^{4/3}$.
For a fixed subhalo mass, we numerically find that the peak cross-section depends on the distance as $\sigma_{\rm FDM}^{\rm peak} \propto d_{\rm sh}^{-2.3}$.
Combining these results, the dependence of the peak cross-section on the subhalo mass and the distance from the host halo center is 
\begin{equation}
    \sigma_{\rm FDM}^{\rm peak}(M_{\rm sh},d_{\rm sh}) \propto M_{\rm sh}^{\frac{4}{3}} d_{\rm sh}^{-2.3}.
\end{equation}
As expected, subhalos near the macro-critical curve and/or those with a larger mass produce larger cross-sections.
Note that we generally ignore the subhalos whose critical curves are larger than 3 arcsec, since such a large critical curve is not observed \cite{2020Sci...369.1347M}.
The smoothing parameter in Eq.~\eqref{ggsl_cs_fdm_model}, $\Delta_{\log_{10} m}$, is an order of $\mathcal{O}(10^{-1})$ and shows no specific trend on the two parameters. 

For a host halo described by a spherical NFW profile with a mass of $M_{\rm hh} = 10^{15}\ M_{\odot}$, and with source and lens redshifts of $z_{\rm s}=2$ and $z_{\rm l}=0.5$, respectively, we numerically determine the coefficients of the peak cross-section and the corresponding peak FDM mass. 
These are expressed in the form of
\begin{eqnarray}
    && \sigma_{\rm FDM}^{\rm peak}(M_{\rm sh},d_{\rm sh}) \simeq 5 \times 10^{-3}\ {\rm arcsec}^{2} \nonumber \\
    && \hspace{18mm} \times \left(\frac{M_{\rm sh}}{10^{11}\ M_{\odot}}\right)^{\frac{4}{3}} \left(\frac{d_{\rm sh}}{20\ {\rm arcsec}}\right)^{-2.3}, \label{sigma_peak_expression} \\
    && m_{\rm peak}(M_{\rm sh}) \simeq 1 \times 10^{-22}\ {\rm eV}/c^{2} \left(\frac{M_{\rm sh}}{10^{11}\ M_{\odot}}\right)^{-1}. \label{m_peak_expression}
\end{eqnarray}
Additionally, we study the empirical relation of the cross-section of a single CDM subhalo as a function of the subhalo mass and the distance to the host halo center, finding 
\begin{eqnarray}
    && \sigma_{\rm CDM}(M_{\rm sh},d_{\rm sh}) \simeq 1\times 10^{-4}\ {\rm arcsec}^{2} \nonumber \\
    && \hspace{18mm}  \times \left(\frac{M_{\rm sh}}{10^{11}\ M_{\odot}}\right)^{2.3} \left(\frac{d_{\rm sh}}{20\ {\rm arcsec}}\right)^{-12}. \label{sigma_cdm_expression}
\end{eqnarray}
The coefficients in Eqs.~\eqref{sigma_peak_expression}, \eqref{m_peak_expression}, and \eqref{sigma_cdm_expression} should depend on the host halo mass and its density distribution, and the redshifts of the source and lens planes, which are fixed in this study.

\subsection{Total GGSL cross-section} \label{subsec:ggsl_fdm_total}
Given the analytical expression for the cross-section of a single CDM/FDM subhalo with different FDM mass, subhalo mass, and distance from the host center, we can now calculate the total cross-section contributed by all subhalos within the host halo.

The total cross-section can be obtained by multiplying the subhalo mass function shown in Sec.~\ref{subsec:subhalo_distribution} to the cross-section of a single halo and integrating over the
subhalo mass and the location of the subhalo.
In the case of the FDM subhalos, the total cross-section is dependent on the FDM mass, 
\begin{eqnarray}
    && \sigma^{\rm tot}_{\rm FDM}(m) = \sum_{i=0}^{N_{\rm sh}} \sigma_{\rm FDM}(m; M_{\rm sh}, d_{\rm sh}) \nonumber \\
    && \hspace{5mm} = \int_{M_{\rm sh, min}}^{M_{\rm sh, max}} dM_{\rm sh} \int_{d_{\rm sh, min}}^{d_{\rm sh, max}} dd_{\rm sh}  \nonumber \\
    && \hspace{15mm} \times \left.\frac{d^{2}N_{\rm sh}}{dM_{\rm sh}dd_{\rm sh}}\right|_{\rm FDM} \sigma_{\rm FDM}(m; M_{\rm sh}, d_{\rm sh}).
\end{eqnarray}
In the case of CDM subhalos, we derive 
\begin{eqnarray}
    &&\sigma^{\rm tot}_{\rm CDM} = \int_{M_{\rm sh, min}}^{M_{\rm sh, max}} dM_{\rm sh} \int_{d_{\rm sh, min}}^{d_{\rm sh, max}} dd_{\rm sh} \nonumber \\
    && \hspace{10mm} \times \left.\frac{d^{2}N_{\rm sh}}{dM_{\rm sh}dd_{\rm sh}}\right|_{\rm CDM} \sigma_{\rm CDM}(M_{\rm sh}, d_{\rm sh}).
\end{eqnarray}

\begin{figure}
    \includegraphics[width=\columnwidth]{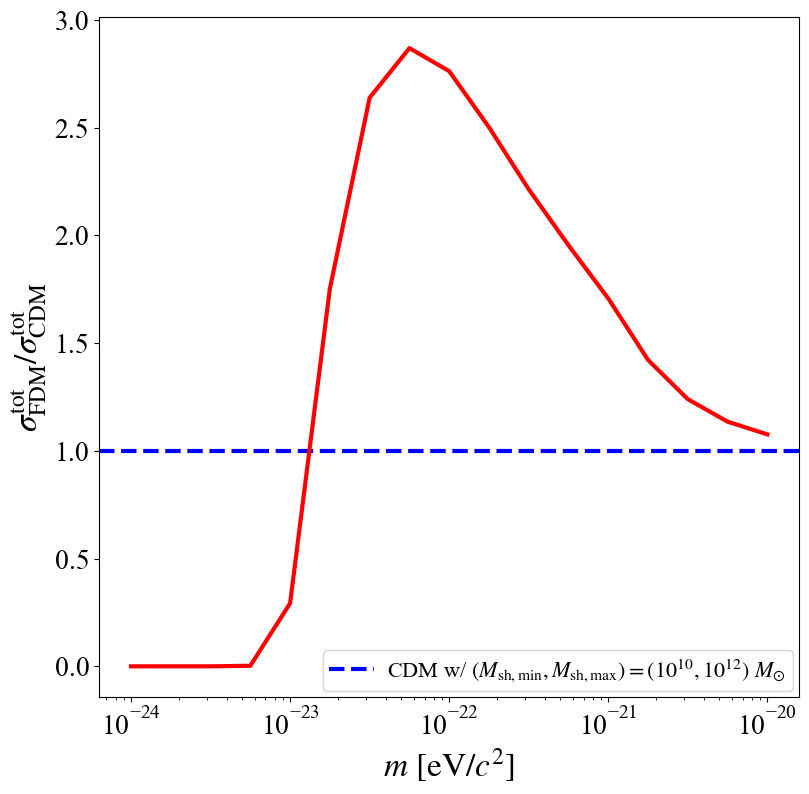}
    \caption{The ratio of the total GGSL cross-sections of the CDM and FDM subhalos as a function of the FDM mass.
    Host halo mass and the redshifts of the source and lens plane are the same as Fig.~\ref{fig:massMap}.
    We set the minimum and maximum subhalo masses to $10^{10}\ M_{\odot}$ and $10^{12}\ M_{\odot}$, respectively.
    The minimum and maximum distances are set to 15 arcsec and 100 arcsec, respectively.
    }
    \label{fig:total_cs}
\end{figure}

In Fig.~\ref{fig:total_cs}, we plot the ratio of the total cross-section between the CDM subhalos and the FDM subhalos as a function of the FDM mass.
Here we set the minimum and maximum subhalo mass to $10^{10}\ M_{\odot}$ and $10^{12}\ M_{\odot}$, respectively.
While the number of subhalos with lower mass $M_{\rm sh} \lesssim 10^{10}\ M_{\odot}$ is large, each cross-section is sufficiently small and they do not largely contribute to the total cross-section.
We set the maximum subhalo mass to $10^{12}\ M_{\odot}$ for the following two reasons.
The first reason is that the number of such high-mass subhalos is small.
The second reason is that such high-mass subhalos would produce large critical curves that we do not observe as stated in Sec.~\ref{subsec:ggsl_fdm_single}.
We set the minimum and maximum distance to the host halo center to $d_{\rm sh, min} = 15\ {\rm arcsec}$ and $d_{\rm sh, max} = 100\ {\rm arcsec}$, respectively.
The minimum distance is set so that the subhalo is located outside the primary critical curve created by the host halo mass distribution.
We find that the cross-section is close to zero with the FDM mass $m \lesssim 10^{-23}\ {\rm eV}/c^{2}$.
Interestingly however, with $m \gtrsim 10^{-23}\ {\rm eV}/c^{2}$, the total cross-section becomes larger than the CDM case.
The peak cross-section is achieved with the FDM mass of $m \simeq 10^{-22}\ {\rm eV}/c^{2}$, where the ratio to the total cross-section from CDM subhalos is close to three.
When the FDM mass is sufficiently large, {\it i.e.}, in the CDM limit, the ratio of the total cross-section is close to one as expected.
While the FDM subhalos can produce a larger cross-section than the CDM subhalos, no FDM mass can provide a sufficiently large cross-section that matches the observation, which is an order of magnitude larger than the CDM case.

\section{Effects of baryons} \label{sec:compare_obs}
The cluster galaxies have baryon components, which we have neglected so far; we now consider the effect of the baryon distribution on the single and total cross-sections.
We include the baryon distribution without modifying the dark matter distribution.
Although this causes the total mass to slightly exceed $10^{15}\ M_{\odot}$, we consider this modification negligible, as the total baryon mass is only about 1\% of the total dark matter mass.
In Sec.~\ref{subsec:baryon}, we briefly explain how we model the baryon profile.
Then we show how baryons affect the GGSL cross-sections in Sec.~\ref{subsec:full_comparison}.

\subsection{Baryon distribution} \label{subsec:baryon}
We assume that the distribution of the baryons (i.e., stars) follows the Hernquist profile \citep{1990ApJ...356..359H},
\begin{equation}
    \rho_{\rm b}(r) = \frac{\rho_{\rm g}}{\frac{r}{r_{\rm g}}\left(1+ \frac{r}{r_{\rm g}}\right)^{3}},
\end{equation}
where $\rho_{\rm g}$ and $r_{\rm g}$ are the characteristic density and radius of the Hernquist profile, respectively.
The characteristic density can be expressed in terms of the characteristic radius and the total stellar mass $M_{\rm s}$,
\begin{equation}
    \rho_{\rm g} = \frac{M_{\rm s}}{2\pi r_{\rm g}^{3}}.
\end{equation}
The characteristic radius is related to the effective radius of the Hernquist profile $r_{\rm e}$ as $r_{\rm g} = 0.551 r_{\rm e}$.
Therefore, the effective radius and the total stellar mass are the two parameters that determine the Hernquist profile.
These two parameters are related to the physical quantities that describe the host subhalo.
The stellar mass can be uniquely determined via the stellar-to-halo mass relation \citep{2020A&A...634A.135G}.
The characteristic radius is related to the virial radius of the (sub)halo as $r_{\rm e} = 0.03 r_{\rm vir}$ \citep{2001ApJ...561...46K, 2012MNRAS.421.3343G}.

In this study, we ignore the back reaction of dark matter distribution due to the presence of the baryons.
This is because we still do not understand how the dark matter density profile, especially in the case of FDM, could be affected by stars in the central region of the galaxy.
We discuss the modification of the dark matter density profile and the resulting cross-section in Sec.~\ref{sec:summary_discussion}.

\subsection{Full comparison} \label{subsec:full_comparison}
\begin{figure*}
    \includegraphics[width=\linewidth]{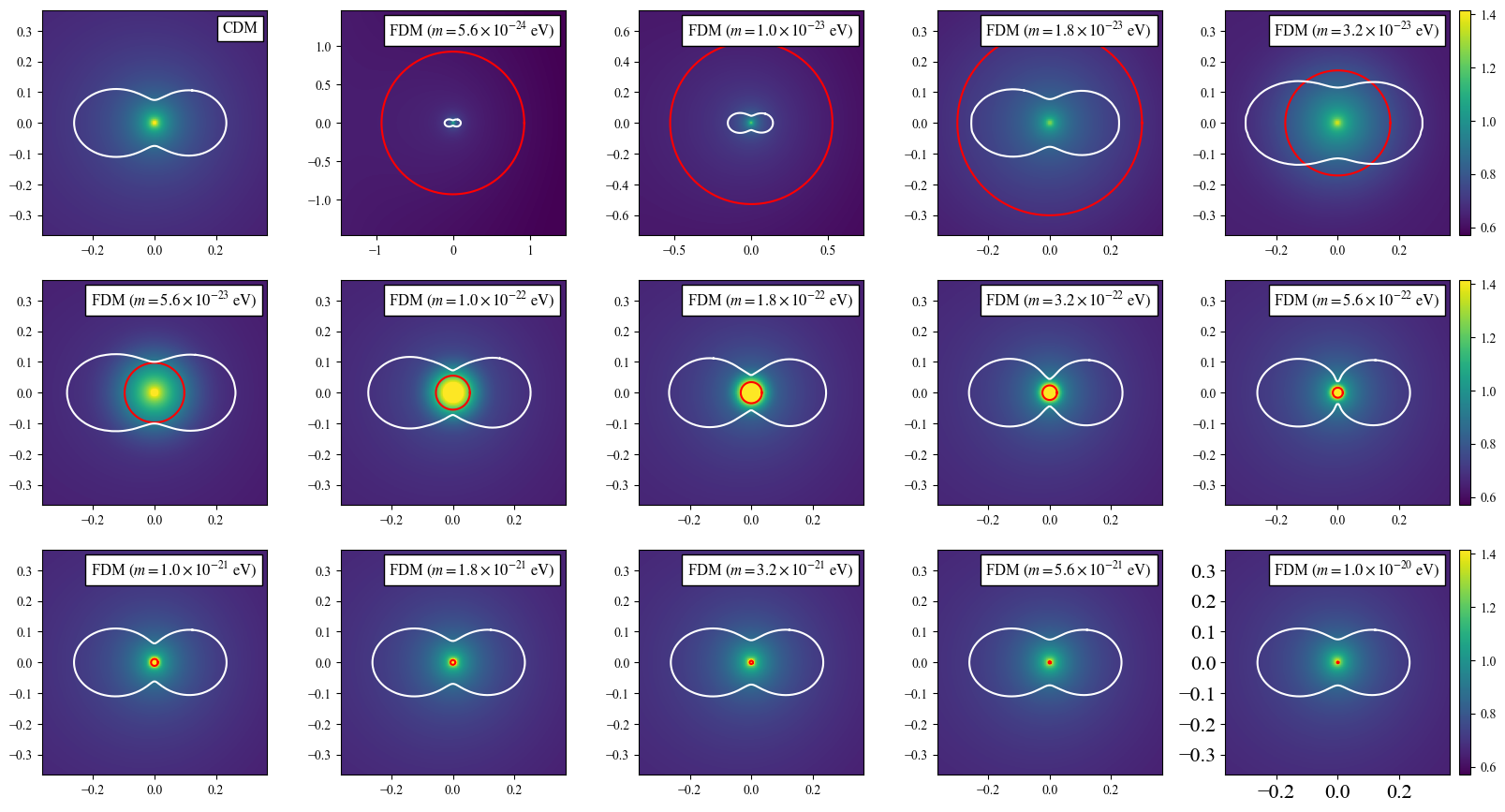}
    \caption{Similar to Fig.~\ref{fig:massMap}, but with the contribution of baryons.
    The baryon distribution is set by the Hernquist profile where we use the stellar-to-halo mass relation studied by  \citet{2020A&A...634A.135G} and the relation $r_{\rm e} = 0.03 r_{\rm vir}$.
    }
    \label{fig:massMap_wb}
\end{figure*}

\begin{figure}
    \includegraphics[width=\columnwidth]{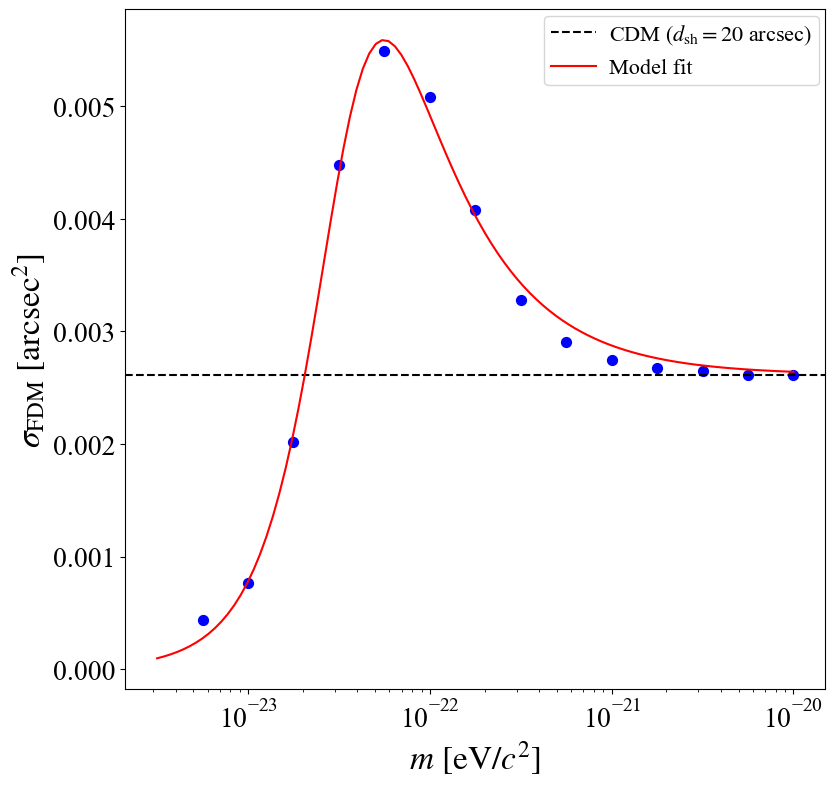}
    \caption{Similar to Fig.~\ref{fig:ggsl_cs}, but with the contribution of baryons.
    The red line shows the fitting result of our analytic model with the fitting parameters being $m_{\rm peak} = 10^{-22.4}\ {\rm eV}/c^{2}$, $\sigma_{\rm FDM}^{\rm peak} = 0.007$, and $\Delta_{\log_{10} m} = 0.31$.
    }
    \label{fig:ggsl_cs_wb}
\end{figure}

Using the baryon distribution above, we explore how the cross-section of each subhalo is changed due to the baryon profile.
In Fig.~\ref{fig:massMap_wb}, we show the convergence maps and critical curves with the same setup as Fig.~\ref{fig:massMap}.
Due to the presence of the baryons, the size of the secondary critical curves around the subhalos becomes larger.
In Fig.~\ref{fig:ggsl_cs_wb}, we plot the corresponding cross-sections.
We can see that the ratio of the peak FDM cross-section to the CDM cross-section is smaller than that without a baryon profile.

While the coefficients and the power-law index are slightly modified, we find that the analytic model shown in Sec.~\ref{subsec:ggsl_fdm_single} can explain the cross-section even with the baryon components, as shown in Fig.~\ref{fig:ggsl_cs_wb}.
The peak cross-section and the peak FDM mass can be expressed by
\begin{eqnarray}
    && \sigma_{\rm FDM}^{\rm peak}(M_{\rm sh},d_{\rm sh}) \simeq 7 \times 10^{-3}\ {\rm arcsec}^{2} \nonumber \\
    && \hspace{18mm} \times \left(\frac{M_{\rm sh}}{10^{11}\ M_{\odot}}\right)^{1.3} \left(\frac{d_{\rm sh}}{20\ {\rm arcsec}}\right)^{-1.7}, \label{sigma_peak_expression_wb} \\
    && m_{\rm peak}(M_{\rm sh}) \simeq 8 \times 10^{-23}\ {\rm eV}/c^{2} \left(\frac{M_{\rm sh}}{10^{11}\ M_{\odot}}\right)^{-1}, \label{m_peak_expression_wb}
\end{eqnarray}
and the CDM cross-section is 
\begin{eqnarray}
    && \sigma_{\rm CDM}(M_{\rm sh},d_{\rm sh}) \simeq 1\times 10^{-3}\ {\rm arcsec}^{2} \nonumber \\
    && \hspace{18mm}  \times \left(\frac{M_{\rm sh}}{10^{11}\ M_{\odot}}\right)^{2.0} \left(\frac{d_{\rm sh}}{20\ {\rm arcsec}}\right)^{-2.3}. \label{sigma_cdm_expression_wb}
\end{eqnarray}
The coefficient of the cross-section becomes larger due to the presence of the baryon profile.
The peak FDM mass is slightly smaller than the case without baryon.
This is because the size of the critical curve increases due to the baryon distribution, and a core of the same size corresponds to a smaller FDM mass.

\begin{figure}
    \includegraphics[width=\columnwidth]{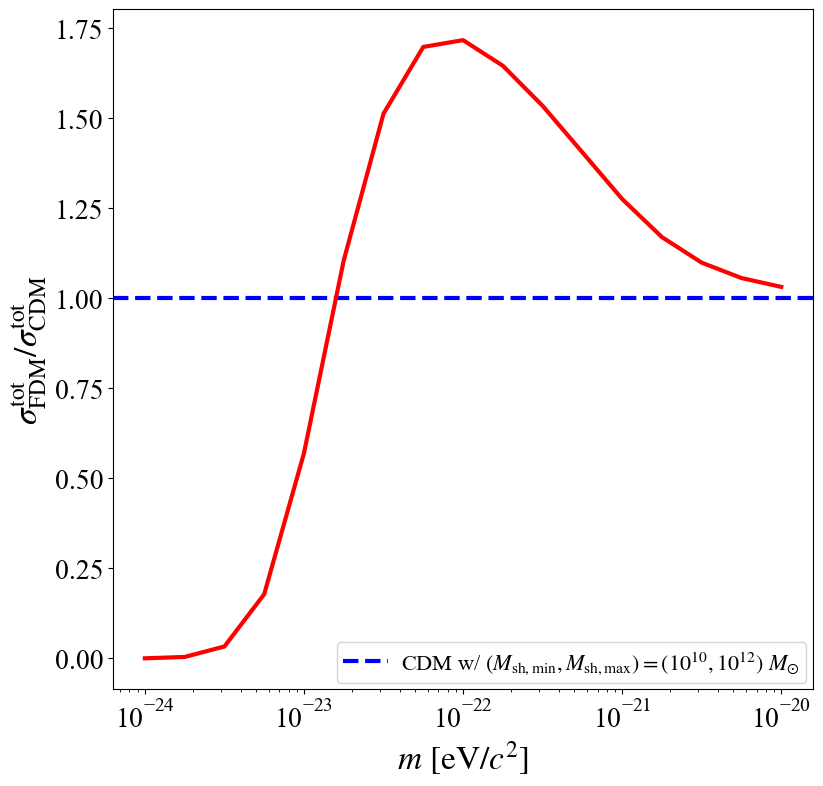}
    \caption{Similar to Fig.~\ref{fig:total_cs}, but with the contribution of baryons.
    }
    \label{fig:total_cs_wb}
\end{figure}

Considering the subhalo mass function, we obtain the total cross-section like in Sec.~\ref{subsec:ggsl_fdm_total}.
In Fig.~\ref{fig:total_cs_wb}, we show the ratio between the total cross-section of the FDM and CDM subhalos as a function of the FDM mass.
Due to the presence of the baryon profile, the ratio is suppressed compared to the case without the baryon profile.
The maximum effect occurs when the FDM mass is about $10^{-22}\ {\rm eV}/c^{2}$ and the ratio is less than two.
Again, since the observations indicate that the total cross-section is about ten times larger than the CDM predictions, it might suggest that the FDM model with any mass cannot explain the observational results.
While it cannot produce a sufficiently large cross-section, the assumptions of spherical host and subhalos, as well as the neglect of changes to dark matter profiles in the presence of baryons, should be revisited in future studies.
These factors could potentially increase the cross-section.

\section{Summary and discussions} \label{sec:summary_discussion}
% Summary
It has been found that there is a discrepancy between the galaxy-galaxy strong lensing cross-section in the observed galaxy clusters with that of the CDM prediction.
One of the possible solutions might be to consider the FDM model, which is the alternative candidate to the CDM.
We have developed an analytic model of the GGSL cross-section of FDM subhalos and compared it with that of the CDM subhalos with the help of numerical simulations.
This study assumes that the host and the subhalos are spherical. The mass distribution of the host halo is described by the NFW profile with a fixed mass of $M_{\rm hh}=10^{15}\ M_{\odot}$, and the redshifts of the source and lens planes are set to $z_{\rm s}=2.0$ and $z_{\rm l}=0.5$, respectively.

We first focus on a single subhalo with mass denoted by $M_{\rm sh}$ located at $d_{\rm sh}$ away from the host halo center.
When the FDM mass is sufficiently small, the central core density is shallow, and there are no critical curves around the subhalos, resulting in the cross-section equal to zero.
As the FDM mass increases, the core radius becomes smaller, and the central core density becomes larger.
When the core radius is almost the same size as the critical curve, the cross-section reaches the maximum.
This can be explained by the fact that the enclosed mass of the soliton core within the core radius is larger than that with the NFW profile.
With a larger FDM mass, the cross-section scales with the mass as $\sigma_{\rm FDM} \propto m^{-1}$ and is closer to the CDM case.
We find that the analytic expression in Eq.~\ref{ggsl_cs_fdm_model} well captures the dependence of the single cross-section on the FDM mass.
We also study the dependence of the two parameters in the analytic expression, the peak FDM mass $m_{\rm peak}$ and the peak cross-section $\sigma_{\rm FDM}^{\rm peak}$, on the subhalo mass $M_{\rm sh}$ and its distance to the host halo center $d_{\rm sh}$.

The total cross-section, which is the sum of the cross-sections from all subhalos, can be obtained by integrating the product of the cross-section of each subhalo and the subhalo mass function over $M_{\rm sh}$ and $d_{\rm sh}$.
In this study, the spatial distribution of the subhalos is assumed to follow the host halo mass distribution.
We find that the FDM mass around $m \simeq 10^{-22}\ {\rm eV}/c^{2}$ yields the highest cross-section, which is several times larger than the total cross-section produced by CDM subhalos.

Since the subhalos with mass $M_{\rm sh} \gtrsim 10^{10}\ M_{\odot}$ contain a sufficient amount of the baryon components, we then consider the effect of the baryon distribution on the single and the total cross-sections.
We assume the baryon distribution follows the Hernquist profile with the total mass obtained from the stellar-to-halo mass relation.
Although the dark matter distribution is likely changed in the presence of stars dominating the mass budget in the central regions of galaxies, we ignore this effect and add the baryon profile to the underlying dark matter. Our choice is motivated by the fact that we still do not know how to describe the modification of the dark matter profile in the presence of baryons analytically.
As expected, the single and total cross-sections become larger due to the presence of the baryons.
The FDM mass dependence on the single cross-section can be expressed similarly to the case without the baryon profile.
The ratio between the peak cross-section of the FDM subhalo and that of the CDM subhalo is smaller compared to the case without the baryons.
This suppression can also be seen in the total cross-section where the peak is about 1.8 times larger than the total CDM cross-section.
Since the observed cross-section is about ten times larger than the CDM prediction, it might indicate that the FDM model with any mass cannot produce a sufficiently large cross-section to match the observations of cluster galaxies.

% Assumption: Ellipticity of host halo
One assumption we should discuss is the sphericity of the host and subhalos.
When we consider the ellipticity of the host and subhalos, we expect that the size of the critical curve would become larger due to the effect of the increased shear.
When the ellipticity is set to less than one, we expect that the peak FDM mass becomes smaller to match the core radius to the size of the enlarged critical curve.
To constrain the FDM mass robustly, further detailed modeling of the ellipticity, as well as the baryon distribution, the proper redshifts of the source and lens planes, and the host halo mass distribution, should be considered.

% Modification of DM density profile
Another simple assumption we should discuss is the ignorance of modifying the dark matter profile when the baryon component is present.
In this study, we add the baryon component because we still do not know how to modify the density profile in the presence of baryons.
In the CDM case, we might obtain it by considering the adiabatic contraction \citep{1986ApJ...301...27B, 2012MNRAS.421.3343G}.
In the FDM model, while the soliton core is expressed with the ground state solution of the SP equation with baryon potential, another condition, such as the core-halo mass relation, is needed to determine the central density of the soliton core, which has not been studied in the literature.
As far as we know, \citet{2020PhRvD.101h3518V} is the only study that conducts a zoom-in simulation of a single FDM halo with baryon physics. 
They find that the soliton core becomes more dense and the core radius smaller.
We expect the peak FDM mass to become smaller than we expect in this study and the peak cross-section would be larger, which might alleviate the discrepancy.

% Future application
By adapting our model, possibly with more realistic extensions, to each galaxy cluster, we expect that it is possible to study the validity of the FDM model more robustly.
Since the cross-section changes as a function of the FDM mass, a further detailed study might be useful to constrain the FDM mass.
Moreover, since the soliton core would affect the strong gravitational lens signal, such as the flux anomaly and the time delay, our model might be applicable to those studies.
We expect that our study is an important step toward uncovering the nature of dark matter.

\begin{acknowledgments}
% We thank the anonymous referee for useful comments.
We thank Laura Leuzzi and Masamune Oguri for useful feedback during this project.
H. K. is supported by JSPS KAKENHI Grant Number JP22J21440.
H. K. thanks the hospitality of INAF OAS Bologna where part of this work was carried out.
It is supported by JSPS KAKENHI Grant Number JP22K21349. 
The research activities described in this paper have been co-funded by the European Union – NextGenerationEU within PRIN 2022 project n.20229YBSAN - Globular clusters in cosmological simulations and in lensed fields: from their birth to the present epoch. MM acknowledges financial support through grants PRIN-MIUR 2020SKSTHZ and INAF  ``The Big-Data era of cluster lensing''.
\end{acknowledgments}

\bibliography{ref} % Produces the bibliography via BibTeX.

\end{document}